\documentclass[11pt]{amsart}

\usepackage[margin=1.2in]{geometry}
\usepackage{times}

\usepackage{amsmath}
\usepackage{amssymb}
\usepackage{amsfonts}
\usepackage{amsthm}
\usepackage{mathrsfs}
\usepackage[all]{xy}
\usepackage{endnotes} 
\usepackage{cite}
\usepackage{graphicx}

\begin{document}

\bibliographystyle{nature}

\begin{titlepage}

\begin{center}
\LARGE{\textbf{Visualization of phase-coherent electron interference in a ballistic graphene Josephson junction }}\\[1.5 cm]
\large{M. T. Allen$^{1}$, O. Shtanko$^{2}$, I. C. Fulga$^{3}$, J. I.-J. Wang$^{2,4}$, D. Nurgaliev$^{1}$, K. Watanabe$^{5}$, T. Taniguchi$^{5}$, A. R. Akhmerov$^{6}$,  P. Jarillo-Herrero$^{2}$, L. S. Levitov$^{2}$, and A. Yacoby$^{1*}$}\\[1.0 cm]

$^{1}$ Department of Physics, Harvard University. Cambridge, MA 02138.

$^{2}$ Department of Physics, Massachusetts Institute of Technology. Cambridge, MA 02139.

$^{3}$ Department of Condensed Matter Physics, Weizmann Institute of Science. Rehovot, Israel.

$^{4}$ School of Engineering and Applied Sciences, Harvard University. Cambridge, MA 02138.

$^{5}$ Environment and Energy Materials Division, National Institute for Materials Science. 1-1 Namiki, Tsukuba, Ibaraki, 305-0044 Japan.

$^{6}$ Kavli Institute of Nanoscience. Delft University of Technology. Lorentzweg 1. 2628 CJ Delft. The Netherlands. \\[1.0 cm]

\textbf{$^{*}$Corresponding author. E-mail: yacoby@physics.harvard.edu}
\end{center}

\end{titlepage}



\begin{quote}
\textbf{Abstract:} Interference of standing waves in electromagnetic resonators forms the basis of many technologies, from telecommunications~\cite{Telecommunications} and spectroscopy~\cite{Spectroscopy} to detection of gravitational waves~\cite{LIGO}.  However, unlike the confinement of light waves in vacuum, the interference of electronic waves in solids is complicated by boundary properties of the crystal, notably leading to electron guiding by atomic-scale potentials at the edges~\cite{GuidedModes1,GuidedModes2,GuidedModes3,Klein7}.  Understanding the microscopic role of boundaries on coherent wave interference is an unresolved question due to the challenge of detecting charge flow with submicron resolution.  Here we employ Fraunhofer interferometry to achieve real-space imaging of cavity modes in a graphene Fabry-P\'erot (FP) resonator, embedded between two superconductors to form a Josephson junction~\cite{Heersche1}.  By directly visualizing current flow using Fourier methods~\cite{Dynes1}, our measurements reveal surprising redistribution of current on and off resonance.  These findings provide direct evidence of separate interference conditions for edge and bulk currents and reveal the ballistic nature of guided edge states.  Beyond equilibrium, our measurements show strong modulation of the multiple Andreev reflection amplitude on an off resonance, a direct measure of the gate-tunable change of cavity transparency. These results demonstrate that, contrary to the common belief, electron interactions with realistic disordered edges facilitate electron wave interference and ballistic transport.\\
\end{quote}

Graphene provides an appealing platform to explore ``electron-optics'' due to the ballistic nature of wavelike carriers and ability to engineer transmission of electronic waves in real space using electrostatic potentials~\cite{RMP1, Klein1, Klein2, Klein4, Klein3, Klein5, Shytov1, Mayorov1}.  In particular, the electronic analog to refractive index is the Fermi energy, which is tunable via electrostatic gating~\cite{Klein1, Williams1}.  Because the gapless spectrum of Dirac materials enables continuous tunability of carrier polarity, positive and negative index of refraction regions can be combined in bipolar structures that form the building blocks of Veselago ``electronic lenses''~\cite{Klein5}, Fabry-P\'erot (FP) interferometers~\cite{Klein1,Klein2, Klein4, Klein3, Klein5, Mayorov1}, and whispering gallery mode cavities ~\cite{Whisper1}.  Electronic analogs to optical interferometers attract attention because relativistic effects such as hyperlensing and phase-coherent Klein transmission provide capabilities beyond conventional optics~\cite{RMP1, Klein1, Klein2, Klein4, Klein3, Klein5, Shytov1, Mayorov1, Gu1}.  Here we investigate the simplest analog to an optical interferometer, the electron FP resonator, which consists of standing electron waves confined between two reflective interfaces~\cite{Liang1, Wu1}. Despite extensive exploration in the momentum domain, in which Fermi momentum is simply tuned with a gate, little information is available about the real-space distribution of current flow due to the challenge of imaging current paths with submicron resolution.  Furthermore, in real devices, atomically sharp potentials at the edges of graphene can confine electron waves into guided edge modes, in analogy to the guiding of light in optical fibers~\cite{GuidedModes1,GuidedModes2,GuidedModes3,Klein7}, as we have demonstrated experimentally in prior work ~\cite{MAllen}.  To investigate the nature of these boundary currents, we measure the interference of standing waves in a graphene Josephson junction and image the real space distribution of supercurrent flow using Fraunhofer interferometry~\cite{Dynes1}.  By visualizing the spatial structure of current-carrying states in the cavity using Fourier methods, our measurements disentangle edge from bulk current flow and highlight the surprising role of the crystal boundaries.\\

In a coherent electron cavity, quantum interference of electron waves replaces classical diffusion as a key feature of electronic transport~\cite{Liang1, Wu1}.  In our system, a pair of superconducting electrodes is coupled to a graphene membrane, defining a ballistic cavity between the two graphene-electrode interfaces.  As the Fermi wavelength in the cavity is tuned with a gate, the quantized energy levels of the cavity are moved on and off resonance with the Fermi energy of the superconducting leads, thus inducing an oscillatory critical current whose period satisfies the FP interference conditions.  Due to the chiral nature of fermions in monolayer and bilayer graphene, at zero magnetic field carrier trajectories with an incidence angle $\theta$ and refraction angle angle $\theta'$ 
produce a contribution to FP fringes in the single-particle transmission probability of the form
\begin{equation}\label{eq:FP}
T(\theta)\sim\frac{|t_1(\theta)|^2 |t_2(\theta)|^2}{|1-r_1(\theta)r_2(\theta)e^{2ik_\parallel L}|^2}
,\quad
k_\parallel =k\cos\theta'
\end{equation}
where $t_{1,2}$ and $r_{1,2}$ are the angle-dependent transmission and reflection amplitudes for the two p-n junctions.
The resulting fringes are dominated by the angles for which both the transmission and reflection are reasonably high (the first harmonic of FP fringes is at its brightest when the product of transmission and reflection coefficients $|t(\theta)|^2|r(\theta)|^2$ takes a maximum value).
In general, a spread of angles for different trajectories in the bulk gives rise to a spread of the FP oscillation periods, 
somewhat reducing the fringe visibility in the net current.  In contrast, no suppression is expected for interference fringes due to edge modes, as discussed in detail later.  \\

We employ proximity induced superconductivity to shed light on the microscopic nature of electron interference in a graphene Josephson resonator~\cite{Nishio1, Jorgensen1, DelftFP, GeimFP}. On a practical level, graphene provides an accessible interface for superconducting electrodes because it is purely a surface material, unlike 2D  electron gases embedded in semiconductor heterostructures. Although graphene is not intrinsically superconducting,  proximity-induced superconductivity can be mediated by phase coherent Andreev reflection at the graphene/superconductor interface.  This process features an electron-hole conversion by the superconducting pair potential that switches both spin and valley to preserve singlet pairing and zero total momentum of the Cooper pair~\cite{Beenakker1}.  In this study, we employ gated mesoscopic Josephson junctions consisting of bilayer graphene suspended between two superconducting Ti/Al electrodes, as well as a graphene device on hBN.  The superconductors serve three roles: (1) they create electrostatic potentials that confine electron waves, serving as electronic analogs to mirrors (2)  superconducting interferometry can extract spatial information on how current flows through the system, and (3) beyond equilibrium, scattering events between the superconductors and graphene (multiple Andreev reflections) depend critically on resonance conditions and reveal how the resonator couples to the outside world.\\

A schematic of a suspended graphene Josephson junction is provided in Figure 1a.  To access the ballistic regime, we developed a new method to isolate the flake from charge disorder in the underlying dielectric by suspending it over the back gate electrode, described in detail in the Supplementary Methods.  This approach combines the high purity of suspended devices with superconductivity enables creation of ballistic waveguides where the mean free path $l_e$ of electrons exceeds channel length $L$.  We note that similar results are also obtained on a gate-defined resonator in monolayer graphene encapsulated in hBN, discussed later, which enables a higher degree of electronic control over the cavity while preserving sample quality.\\

The superconducting leads serve not only as electronic probes but also induce a resonant electron cavity in the scaling limit $l_e>L$ (Fig. 1b)~\cite{Liang1, Wu1}.  The graphene in the immediate vicinity of the Ti/Al contact is n-doped by charge transfer ~\cite{Blake1}, forming an intrinsic n-n or n-p junction near the interface when the graphene has electron or hole carriers, respectively.  We exploit contact induced doping to define the resonator because it is scalable to ultrashort channel lengths, provides electrostatic barriers that are sharp compared to the electron wavelength, and is less complex than gate-defined methods~\cite{Young1, Grushina1, Rickhaus1, Allen1}.  Analogous to an optical Fabry-P\'erot cavity, the n-p junctions serve as the electronic counterparts to mirrors while the ballistic graphene channel serves as an electron waveguide. The Fermi wavelength $\lambda_F$ of electrons in the cavity is directly tunable with a gate electrode which controls the carrier density $n$.\\

Fabry-P\'erot (FP) resonances in ballistic junctions arise due to reflection from p-n junctions formed near superconducting leads when carrier polarity in the graphene region is opposite to the polarity of contact doping.  Figure 1b shows a plot of the normal resistance $R_n$, obtained by sweeping the gate voltage $V_b$ at a fixed bias exceeding $I_c$.  We observe well-resolved resistance oscillations at small positive carrier densities ($V_b<0$) when n-p-n junction formation is favored and monotonic behavior when doping is unipolar.  The dips in $R_n$ coincide with carrier densities satisfying the constructive interference condition $2d=m\lambda_F$ for electron waves in a resonator, where $d$ is the effective cavity length and $m$ is an integer.  Sweeping the gate voltage changes the Fermi energy in the graphene and hence the Fermi wavenumber, given by $k_F=2\pi/\lambda_F=\sqrt{\pi n}$ for a 2D Fermi disk with fourfold degeneracy.  The correspondence to FP interference conditions can be seen more clearly in Supplementary Figure S1, which shows that $R_n$ is periodic as a function of $2d/\lambda_F$. Reproducibility of the oscillation period is demonstrated in three devices of 500 nm length (Fig. S1), while shorter 
junctions exhibit larger periods as expected.  Quantum confinement between the cavity ``mirrors'' gives rise to discrete energy levels with spacing $h v_F/2d$, where $v_F=\hbar k_F/m^*$ is the Fermi velocity and and $m^*$ is the effective electron mass in bilayer graphene.  We evaluate this energy scale to be of the order 1 meV using the height of FP diamonds, as measured using voltage bias spectroscopy (Fig. S1).\\

The interplay between cavity resonances and supercurrent is evident from a resistance colormap as a function of $I_{DC}$ and $V_b$ (Fig. 1c-d) showing critical current oscillations whose period satisfies FP interference conditions, consistent with supercurrent propagation via ballistic charge carriers~\cite{Rittenhouse1}.  As $\lambda_F$ in the cavity is tuned with the gate, the quantum levels of the cavity are moved \textit{on} or \textit{off} resonance with the Fermi energy of the superconducting leads, thus inducing a oscillating critical current periodic in $\sqrt{n}$ for bilayer graphene.  This phenomena is observed in two independent systems: (1) suspended bilayer graphene resonators defined by contact-induced doping (Fig. 1d) and (2) a gate-defined resonator in monolayer graphene on hBN (Fig. 1e and Supplementary Fig. S2), both of which exhibit similar behavior.  In total, five suspended bilayer devices are studied with a lithographic distance $L$ between superconducting contacts of 350 to 500 nm and contact width $W$ of 1.5 to 3.2 $\mu$m, in addition to one gate-defined monolayer device with cavity dimensions of $L=100$ nm and $W=2.7$  $\mu$m (see Supplementary Methods).    Figure 1e displays critical current modulations in a gate-defined monolayer resonator whose oscillations are periodic in $n$, in agreement with a monolayer FP model  for cavity length $\sim 100$ nm.    \\

Next we employ superconducting interferometry as a tool to spatially resolve  optics-like phenomena associated with electron waves confined within a ballistic graphene Josephson junction.  Unlike experiments in 1D systems~\cite{Nishio1, Jorgensen1, PJH1}, one can thread flux through the junction and explore the rich interplay between magnetic interference effects and cavity transmission.  Upon application of a magnetic field $B$, a flux $\Phi$ penetrates the junction area and induces a superconducting phase difference $\Delta \phi(x)= 2\pi \Phi x/ \Phi_0 W$ parallel to the graphene/contact interface, where $\Phi_0=h/2e$ is the flux quantum, $h$ is Planck's constant, and $e$ is the elementary charge.  When a flux penetrates the junction area, the critical current $I_c(B)$ exhibits oscillations in magnetic field given by:
\begin{equation}
\label{Fraunhofer_eqn}
I_c(B)= \left|\int_{-W/2}^{W/2} J(x)\cdot e^{2\pi iLBx/\Phi_0} dx \right|
\end{equation}
where $L$ is the distance between superconducting electrodes (Fig. 1)~\cite{TinkhamBook, Dynes1}.  This integral expression applies in the wide junction limit, relevant for our system, where $L\ll W$ and the current density is only a function of one coordinate.  Because the critical current $I_c(B)$ equals magnitude of the complex Fourier transform of the real-space supercurrent distribution $J(x)$, the shape of the interference pattern is determined directly by the spatial distribution of supercurrent across the sample~\cite{Dynes1, DasSarma1}.\\

To visualize current flow associated with interfering electron waves in graphene, we measure supercurrent modulations in $B$ field that arise from a Fraunhofer diffraction. Figure 2a is a color map of critical current $I_c$ as a function of gate voltage and magnetic field.  Each pixel is obtained by measuring the DC voltage $V_{sd}$ across the junction as a function as a function of applied DC current bias $I_{DC}$ and extracting the maximum derivative $dV_{sd}/dI_{DC}$.   In a conventional graphene Josephson junction with uniform current density, the normalized critical current
$I_c(B)/I_c(0) = |\sin (\pi \Phi/ \Phi_0) / (\pi \Phi / \Phi_0) |$ is described by Fraunhofer diffraction and should be independent of gate voltage.  Our results exhibit a striking departure from this picture and feature nodes in $I_c(B)/I_c(0)$ as a function of both $V_b$ and $B$ (as shown in Supplementary Fig. S3).  Figure 2a and Fig. S3 display the different behavior of $I_c$ versus $B$ on and off resonance, where the red and green dotted lines indicate gate voltages corresponding to on and off resonance conditions, respectively.  (Reproducibility of this phenomenon in additional samples is shown in Supplementary Fig. S4.)  Using Eq.\eqref{Fraunhofer_eqn}, one can extract an effective spatial distribution of the supercurrent $J(x)$ by taking the inverse Fourier transform of the above $I_c(B)$ line plots with the technique of Dynes and Fulton~\cite{Dynes1} (see Supplementary Methods).  As revealed in Fig. 2b, the normalized spatial distribution features bulk-dominated current flow on resonance and an enhanced edge current contribution off resonance.  \\

Inspired by the relation between the spatial current distribution $J(x)$ and critical current $I_c(B)$ in Eq. \eqref{Fraunhofer_eqn}, we directly model the spatial distribution of current paths for bilayer graphene in the FP regime (Fig. 2c-e).  These calculations take into account guided edge modes due to band-bending at the crystal boundaries, which have been experimentally observed in Ref. \cite{MAllen}.  This electron guiding effect can be quantified by an edge potential, which is capable of confining carriers to edge-defined `waveguides' in analogy to the confinement of photons in fiber optic cables.  Energies of these edge states lie outside the bulk continuum (Fig. 2c), which ensures an evanescent-wave decay of carrier states into the bulk. The resulting states are effectively one-dimensional, propagating as plane waves along the graphene edges. Applying the FP quantization condition in the p-n-p region leads to a sequence of FP maxima positioned at  $k_n=\pi n/L$, where $n$ is an integer and $L$ represents distance between superconducting contacts. These quasi-1D states guided along the edge feature head-on transmission and reflection and hence should produce much stronger FP fringes than the bulk states.\\

As shown in the theoretical dispersion in Fig. 2c, the interference conditions in the bulk and at the edge should not coincide due to the difference in the carrier dispersion at the edge and in the bulk  as well as due to the angle-dependence of the FP period for the latter carriers. Hence a gradual increase of doping will trigger repeated switching between the bulk-dominated and edge-dominated regimes, with the current distribution switching from an approximately uniform to edge like, accordingly.
Qualitatively, this would be manifested in the dependence of measured critical current on applied magnetic field, switching between Fraunhofer and more SQUID-like behavior (Fig. 2a).\\

To quantify these phenomena, we model FP resonances using the approach described in Ref.\cite{MAllen}.  Assuming that the edge potential is sufficiently short-ranged, we
approximate it with a delta function.  We obtain the density of persistent current along the edge (chosen to be along $y$ axis) from the exact Green's function $G$  in a mixed coordinate-momentum representation:
\begin{equation}
\label{current_formula}
j(\epsilon,x) = -\frac{\pi}{\rm Im}\sum_{k_y,x'=x} \Bigl[G(\epsilon,x,x',k_y)J_y\Bigl]
\end{equation}
where $J_y$ is the operator for current along the edge (see Eq. (8-9) in supplement of Ref.\cite{MAllen}).
The sum in Eq.\eqref{current_formula} runs over the values  $k_n=\pi n/L$ for one sign of $n$.
Each term in Eq.\eqref{current_formula} has poles corresponding to bounded states for the momentum value $k_n$, each of which corresponds a to current maximum at the edge (see Fig. 2e).  The predicted spatially resolved current density $j(x)$ across the sample as function of energy is shown in Figure 2e. To translate this into an experimentally observable quantity, we model Fraunhofer interference pattern $I_c(B)$ using the theoretical amplitude and spatial distribution of edge modes (see Supplementary Materials).  This result, plotted in Fig. 2d, captures the key features of the data, namely the redistribution of current on and off resonance as well as the suppression of side lobes' intensity on resonance. Thus, the measurements are consistent with a model that features separate FP interference of guided-wave edge currents, in parallel to interference of bulk modes.  This further suggests that the quasi-1D edge currents previously observed ~\cite{MAllen} have ballistic character.  Despite its simplified nature, which neglects disorder and finite temperature effects, our model captures the essential features of the measurements.  While the edge potential featured in this simulation accommodates a single edge channel, we note that the number of guided modes may exceed one for stronger potentials. In this case, each mode would contribute independently to the interference pattern, giving rise to fringes with complicated multi-period structure at the edge.\\

We employ yet another property of superconductor-normal-superconductor (SNS) systems to gain insight into the coupling between the cavity modes with the superconducting reservoirs.  Because the phenomenon of multiple Andreev reflection (MAR) is known to be extremely sensitive to the coupling between electrons in the normal metal and the superconductor, we use voltage bias spectroscopy to map out the interplay between MAR oscillation amplitude and cavity transmission (Fig. 3a,b).  The millielectronvolt energy scale associated with FP interference substantially exceeds the Al superconducting gap $\Delta$, allowing one to study the system close to equilibrium conditions for the resonator.  A colormap of resistance $R_n$ as a function of applied voltage bias $V_{DC}$ and gate voltage $V_b$ shows modulations due to FP interference (Fig. 3b). Well defined MAR peaks appear at $2\Delta, \Delta$, and $2\Delta/3$ when the density is tuned off resonance, while MAR is completely suppressed on resonance, as visible in line cuts of resistance on and off resonance in Fig. 3c (additional data sets are provided in Supplementary Figs. S5-S6).   It is notable that the amplitude of the multiple Andreev reflections depends strongly on cavity resonance conditions, thereby providing a direct measure of the tunable coupling between the resonator and the outside world.  \\

The change in visibility of MAR on and off FP resonances is most naturally explained by changes in the distribution of transmission eigenvalues, which can be understood using the following model.  Because supercurrent is predominately transmitted by bulk modes, as indicated by the Fraunhofer interferometry data (Fig. 2), we simplify our analysis by focusing on resonances of bulk states.  The magnitude of multiple Andreev reflection peaks is small for modes with high transmission probability due to the absent suppression of higher order scattering processes\cite{Averin95}. 
In a FP cavity a larger fraction of the current is carried by highly transmitting modes when the cavity is tuned to the resonant wave length. 
In a short junction different modes contribute independently to the current, producing the observed multiple Andreev reflection pattern.  In our junction,  $\xi = \hbar v_F / \Delta \approx 450-700$ nm, while the junction size is $\approx 350$ nm, so we expect the short junction limit to qualitatively hold.\\

In order to compare this model to the observed experimental data, we have modeled the current through the junction as a sum of contributions of the modes with high ($\sim 0.9$), medium ($\sim 0.6$), and low transmission ($\sim 0.3$) coefficient (Fig. S7). 
This separation was chosen to avoid overfitting, while keeping the qualitative features of I-V relationships with different transparencies. We then approximate
\begin{equation}\label{eq:Gv}
 I(V) = \sum_n \rho(T_n) \cdot I(V, T_n),
\end{equation}
with $T_n$ the transmission probability in various channels, $\rho$ the density of transmission eigenvalues, and $I(V, T)$ the contribution of a single mode with transmission probability $T$ to the total current, calculated in the short junction limit following Ref.\cite{Averin95}.  Fitting the model to the measured conductance curves on and off resonance (Fig. 3c,d and Supplementary Fig. S7) shows that the junction transparency is increased on resonance and suggests good qualitative agreement between this theoretical interpretation and the experiment.\\

We obtain the estimated contributions of each $T_n$ by fitting the measured I-V traces using the Eq.~\eqref{eq:Gv} constrained by the condition $\rho(T_n) > 0$. 
The fits show no systematic error, and increasing the number of $T_n$ leads to noisier fits, indicating overfitting. 
The coefficient $\rho$ corresponding to large transmissions increase, while the ones corresponding to low transmissions decrease whenever the system is on resonance, at values of the back gate voltage where the normal state conductance is also peaked (see Supplementary Fig. S7). 
The normal state conductance estimated using our model $G_N=g_0\sum_n T_n\rho(T_n)$ is smaller than the measured one for all back gate voltages, which may be due to deviations from the short junction theory, or the nonlinear behavior of the p-n junctions.\\

In summary, we utilize different aspects of proximity-induced superconductivity, particularly Fraunhofer interferometry and Andreev scattering, as new tools to resolve  optics-like phenomena associated with electron waves confined within a ballistic graphene Josephson junction.  This enables real-space visualization of cavity modes in a graphene FP resonator, which reveals surprising redistribution of current on and off resonance and provides direct evidence of the ballistic nature of guided edge currents.  These results constitute a strong departure from conventional Josephson behavior in graphene and motivate further exploration of new effects at the intersection of superconductivity and optics-like phenomena.\\

\section*{Acknowledgments}
The authors thank O. Dial, B. Halperin, V. Manucharyan, and J. Sau for helpful discussions. This work is supported by the Center for Integrated Quantum Materials (CIQM) under NSF award 1231319 (LSL and OS) and the U.S. DOE Office of Basic Energy Sciences, Division of Materials Sciences and Engineering under award DE-SC0001819 (PJH, JIW, MTA, AY). Nanofabrication was performed at the Harvard Center for Nanoscale Systems (CNS), a member of the National Nanotechnology Infrastructure Network (NNIN) supported by NSF award ECS-0335765.  AA was supported by the Foundation for Fundamental Research on
Matter (FOM), the Netherlands Organization for Scientific Research
(NWO/OCW). ICF was supported by the European Research Council under
the European Union's Seventh Framework Programme (FP7/2007-2013) / ERC
Project MUNATOP, the US-Israel Binational Science Foundation, and the
Minerva Foundation for support.

\section*{Figure legends}

\textbf{Figure 1.}$ \quad$\textbf{Interplay between superconductivity and the Fabry-P\'erot interference in a  ballistic graphene Josephson junction}. \textbf{(a)} Gated mesoscopic Josephson junction consisting of bilayer graphene suspended between two superconducting Ti/Al electrodes.  $L$ is the lithographic distance between contacts and $W$ is the junction width.  In the presence of magnetic field, a flux threads the junction area.  A current bias is applied between the electrodes and the voltage drop across the device is recorded.  A voltage applied to the back gate electrode $V_b$ tunes the Fermi wavelength $\lambda_F$ in the cavity.  \textbf{(b)} Plot of the normal resistance, obtained by sweeping the gate voltage $V_b$ at a fixed bias exceeding $I_c$.  Data sets in panels (b-d) are from device \textit{B1}. \textit{Left inset:} Charge transfer at the boundaries of the superconducting electrodes leads to intrinsic $n$-doped regions near the contacts, forming an electronic resonator when the bulk is tuned to hole doping. Dips in resistance appear when constructive interferences conditions in the cavity are satisfied, $2L=m\lambda_F$.  \textit{Right inset:} When the bulk is tuned to electron doping, standing waves are not formed, leading to monotonic resistance.  \textbf{(c, d)} Plots of resistance as a function of DC current bias and back gate voltage. The critical current $I_c$ oscillates with a period that satisfies the Fabry-P\'erot (FP) interference conditions, consistent with supercurrent propagation via ballistic charge carriers. \textbf{(e)}Differential resistance of a gate-defined FP resonator in monolayer graphene on hBN (device \textit{M1}), as a function of top gate voltage and DC bias current when the back gate voltage is held fixed at -1.75 V. The critical current, defined by the width of zero resistance region along the current axis, oscillates with the same periodicity as normal state resistance, in agreement with a FP model  for cavity length $~\sim 100$ nm. \\

\textbf{Figure 2.}$ \quad$\textbf{Spatially resolved supercurrent imaging in a ballistic graphene cavity}. \textbf{(a)} Plot of critical current $I_c$ as a function of back gate voltage $V_b$ and applied magnetic field $B$.  Each pixel was obtained by measuring the DC voltage $V_{sd}$ across the junction as a function as a function of DC current bias $I_{DC}$ and extracting the maximum derivative $dV_{sd}/dI_{DC}$.  Red and green dotted lines indicate on and off resonance conditions, respectively.  Data was collected from device \textit{B2}.  \textbf{(b)}  Real-space normalized supercurrent density distribution $J(x)/J_{max}(x)$ extracted from the $I_c(B)$ data in (a) using Fourier techniques (see Supplement for details). \textbf{(c)} Spectrum of bilayer graphene with small edge potential, for which one edge mode dominates.  \textbf{(d)} Theoretical plot of critical current $I_c$ as a function of barrier energy and applied magnetic field in presence of edge modes. Bulk and edge currents produce distinct FP patterns due to different dispersion laws and angle dependent transmission of bulk modes.  \textbf{(e)}  Theoretical calculation of spatially resolved current density across the sample as function of energy. Here $p_0 = \lambda m^*/2\hbar$, $E_0 = p_0^2/2m^*$ and $x_0 = \hbar/p_0$ with $m^*$=0.04 $m_e$ (BLG band mass) and delta function potential strength $\lambda = 0.5$ eV$\cdot$nm (see Ref.\cite{MAllen}). Energies corresponding to quantized momenta are represented by horizontal red lines. \\

\textbf{Figure 3.}$ \quad$\textbf{Interplay between multiple Andreev reflections and cavity transmission}. \textbf{(a)} Schematic illustration of the mechanism of multiple Andreev reflection in a graphene Josephson junction for voltage bias $eV=2\Delta/3$. \textbf{(b)} \textit{Right panel}: A colormap of resistance $R_n$ as a function of applied voltage bias $V_{DC}$ and gate voltage $V_b$ shows modulations due to Fabry-P\'erot interference. \textit{Left panel}: Derivative plot $dR_n/dV_{DC}$ for the data on the right.  Data sets in panels (c,d) are from device \textit{B3}. \textbf{(c)}  Line cuts of resistance versus DC voltage bias \textit{on} ($V_b$=0.3 V, red curve) and \textit{off} ($V_b$=0.14 V, blue curve) resonance.  Well defined MAR peaks appear at $2\Delta, \Delta$, and $2\Delta/3$ when the density is tuned off resonance, while MAR is completely suppressed on resonance. \textbf{(d)}Theoretically obtained conductance profiles in the short junction limit, as a function of applied bias voltage. The curve corresponding to high transmission, $G_{\rm high}$ (red) is computed for a single mode with transmission $0.9$. The low transmission curve (blue) is obtained for $4$ modes with transmission $0.6$. Lower transparencies lead to the formation of conductance resonances at bias voltages corresponding to $2\Delta/3$, $\Delta$, and $2\Delta$.

\newpage
\textbf{Figure 1}
\begin{figure}[!h]
\includegraphics[width=160mm]{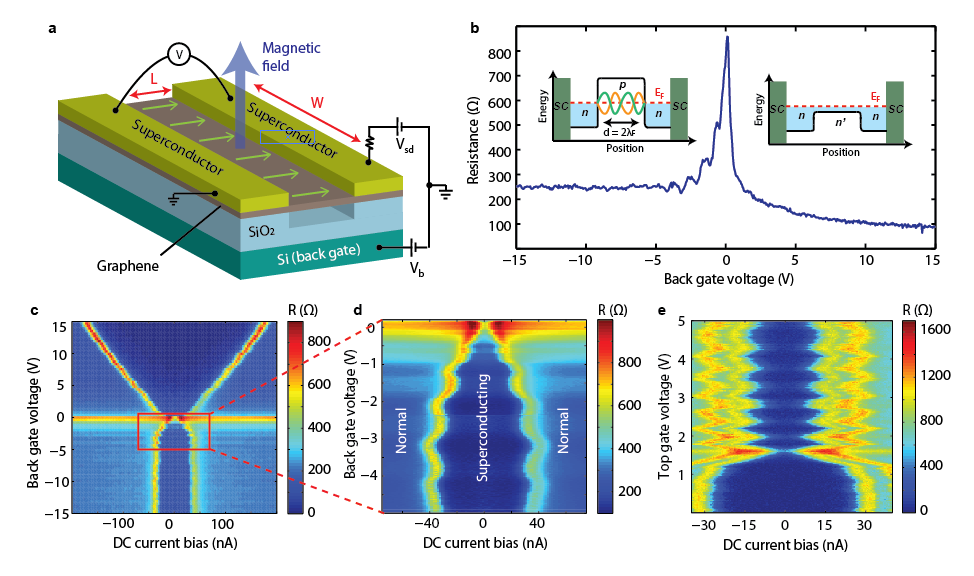}
\end{figure}

\newpage
\textbf{Figure 2}
\begin{figure}[!h]
\includegraphics[width=160mm]{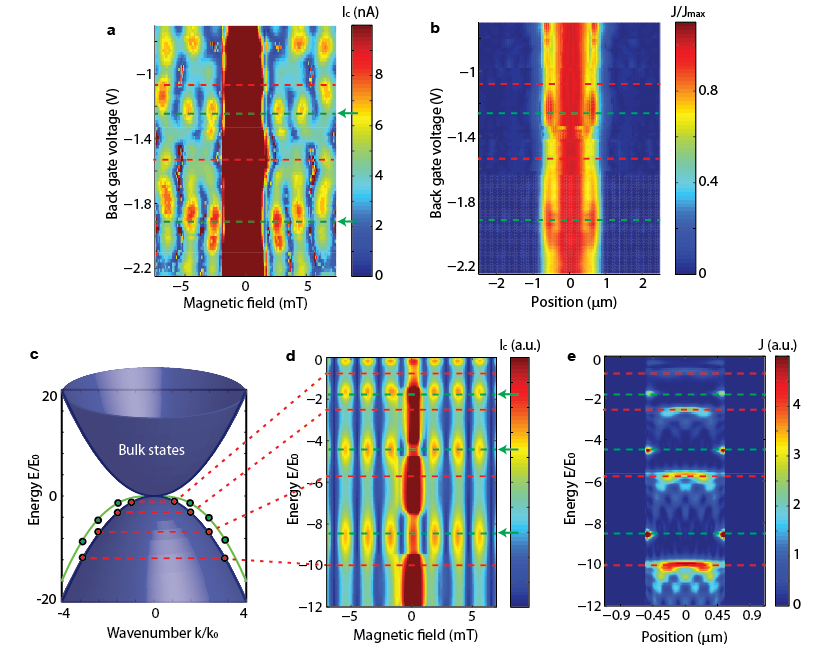}
\end{figure}

\newpage
\textbf{Figure 3}
\begin{figure}[!h]
\includegraphics[width=160mm]{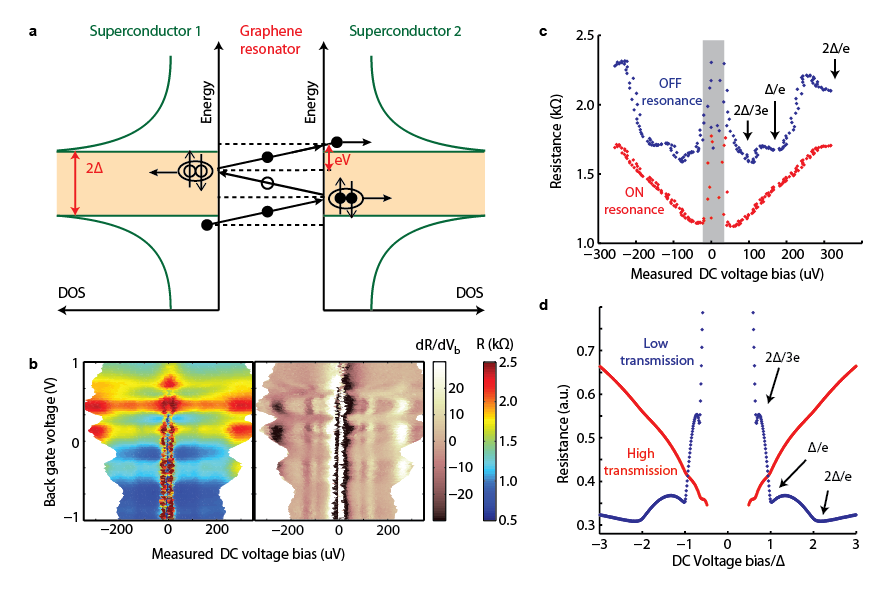}
\end{figure}


\begin{thebibliography}{10}
\expandafter\ifx\csname url\endcsname\relax
  \def\url#1{\texttt{#1}}\fi
\expandafter\ifx\csname urlprefix\endcsname\relax\def\urlprefix{URL }\fi
\providecommand{\bibinfo}[2]{#2}
\providecommand{\eprint}[2][]{\url{#2}}

\bibitem{Telecommunications}
\bibinfo{author}{Spisser, A.} \emph{et~al.}
\newblock \bibinfo{title}{Highly selective and widely tunable 1.55 um
  {I}n{P}/air-gap micromachined {F}abry-{P}erot filter for optical
  communications}.
\newblock \emph{\bibinfo{journal}{Photonics Technology Letters, IEEE}}
  \textbf{\bibinfo{volume}{10}}, \bibinfo{pages}{1259--1261}
  (\bibinfo{year}{1998}).

\bibitem{Spectroscopy}
\bibinfo{author}{Balle, T.~J.} \& \bibinfo{author}{Flygare, W.~H.}
\newblock \bibinfo{title}{Fabry-{P}erot cavity pulsed {F}ourier transform
  microwave spectrometer with a pulsed nozzle particle source}.
\newblock \emph{\bibinfo{journal}{Review of Scientific Instruments}}
  \textbf{\bibinfo{volume}{52}}, \bibinfo{pages}{33--45}
  (\bibinfo{year}{1981}).

\bibitem{LIGO}
\bibinfo{author}{Abramovici, A.} \emph{et~al.}
\newblock \bibinfo{title}{L{I}{G}{O}: The laser interferometer
  gravitational-wave observatory}.
\newblock \emph{\bibinfo{journal}{Science}} \textbf{\bibinfo{volume}{256}},
  \bibinfo{pages}{325--333} (\bibinfo{year}{1992}).

\bibitem{GuidedModes1}
\bibinfo{author}{Pereira, J.~M.}, \bibinfo{author}{Mlinar, V.},
  \bibinfo{author}{Peeters, F.~M.} \& \bibinfo{author}{Vasilopoulos, P.}
\newblock \bibinfo{title}{Confined states and direction-dependent transmission
  in graphene quantum wells}.
\newblock \emph{\bibinfo{journal}{Phys. Rev. B}} \textbf{\bibinfo{volume}{74}},
  \bibinfo{pages}{045424} (\bibinfo{year}{2006}).

\bibitem{GuidedModes2}
\bibinfo{author}{Zhang, F.-M.}, \bibinfo{author}{He, Y.} \&
  \bibinfo{author}{Chen, X.}
\newblock \bibinfo{title}{Guided modes in graphene waveguides}.
\newblock \emph{\bibinfo{journal}{Applied Physics Letters}}
  \textbf{\bibinfo{volume}{94}} (\bibinfo{year}{2009}).

\bibitem{GuidedModes3}
\bibinfo{author}{Hartmann, R.~R.}, \bibinfo{author}{Robinson, N.~J.} \&
  \bibinfo{author}{Portnoi, M.~E.}
\newblock \bibinfo{title}{Smooth electron waveguides in graphene}.
\newblock \emph{\bibinfo{journal}{Phys. Rev. B}} \textbf{\bibinfo{volume}{81}},
  \bibinfo{pages}{245431} (\bibinfo{year}{2010}).

\bibitem{Klein7}
\bibinfo{author}{Williams, J.~R.}, \bibinfo{author}{Low, T.},
  \bibinfo{author}{Lundstrom, M.~S.} \& \bibinfo{author}{Marcus, C.~M.}
\newblock \bibinfo{title}{Gate-controlled guiding of electrons in graphene}.
\newblock \emph{\bibinfo{journal}{Nat. Nanotech.}}
  \textbf{\bibinfo{volume}{6}}, \bibinfo{pages}{222--225}
  (\bibinfo{year}{2011}).

\bibitem{Heersche1}
\bibinfo{author}{Heersche, H.~B.}, \bibinfo{author}{Jarillo-Herrero, P.},
  \bibinfo{author}{Oostinga, J.~B.}, \bibinfo{author}{Vandersypen, L. M.~K.} \&
  \bibinfo{author}{Morpurgo, A.~F.}
\newblock \bibinfo{title}{Bipolar supercurrent in graphene}.
\newblock \emph{\bibinfo{journal}{Nature}} \textbf{\bibinfo{volume}{446}},
  \bibinfo{pages}{56--59} (\bibinfo{year}{2007}).

\bibitem{Dynes1}
\bibinfo{author}{Dynes, R.~C.} \& \bibinfo{author}{Fulton, T.~A.}
\newblock \bibinfo{title}{Supercurrent density distribution in {J}osephson
  junctions}.
\newblock \emph{\bibinfo{journal}{Phys. Rev. B}} \textbf{\bibinfo{volume}{3}},
  \bibinfo{pages}{3015} (\bibinfo{year}{1971}).

\bibitem{RMP1}
\bibinfo{author}{Castro~Neto, A.~H.}, \bibinfo{author}{Guinea, F.},
  \bibinfo{author}{Peres, N. M.~R.}, \bibinfo{author}{Novoselov, K.~S.} \&
  \bibinfo{author}{Geim, A.~K.}
\newblock \bibinfo{title}{The electronic properties of graphene}.
\newblock \emph{\bibinfo{journal}{Rev. Mod. Phys.}}
  \textbf{\bibinfo{volume}{81}}, \bibinfo{pages}{109--162}
  (\bibinfo{year}{2009}).

\bibitem{Klein1}
\bibinfo{author}{Katsnelson, M.~I.}, \bibinfo{author}{Novoselov, K.~S.} \&
  \bibinfo{author}{Geim, A.~K.}
\newblock \bibinfo{title}{Chiral tunnelling and the klein paradox in graphene}.
\newblock \emph{\bibinfo{journal}{Nat. Phys.}} \textbf{\bibinfo{volume}{2}},
  \bibinfo{pages}{620--625} (\bibinfo{year}{2006}).

\bibitem{Klein2}
\bibinfo{author}{Young, A.~F.} \& \bibinfo{author}{Kim, P.}
\newblock \bibinfo{title}{Quantum interference and {K}lein tunnelling in
  graphene heterojunctions}.
\newblock \emph{\bibinfo{journal}{Nat. Phys.}} \textbf{\bibinfo{volume}{5}},
  \bibinfo{pages}{222--226} (\bibinfo{year}{2009}).

\bibitem{Klein4}
\bibinfo{author}{Campos, L.} \emph{et~al.}
\newblock \bibinfo{title}{Quantum and classical confinement of resonant states
  in a trilayer graphene {F}abry-{P}erot interferometer}.
\newblock \emph{\bibinfo{journal}{Nat. Commun.}} \textbf{\bibinfo{volume}{3}},
  \bibinfo{pages}{1239} (\bibinfo{year}{2012}).

\bibitem{Klein3}
\bibinfo{author}{Varlet, A.} \emph{et~al.}
\newblock \bibinfo{title}{Fabry-{P}\'erot interference in gapped bilayer
  graphene with broken anti-{K}lein tunneling}.
\newblock \emph{\bibinfo{journal}{Phys. Rev. Lett.}}
  \textbf{\bibinfo{volume}{113}}, \bibinfo{pages}{116601}
  (\bibinfo{year}{2014}).

\bibitem{Klein5}
\bibinfo{author}{Cheianov, V.~V.}, \bibinfo{author}{Fal'ko, V.} \&
  \bibinfo{author}{Altshuler, B.~L.}
\newblock \bibinfo{title}{The focusing of electron flow and a {V}eselago lens
  in graphene p-n junctions}.
\newblock \emph{\bibinfo{journal}{Science}} \textbf{\bibinfo{volume}{315}},
  \bibinfo{pages}{1252--1255} (\bibinfo{year}{2007}).

\bibitem{Shytov1}
\bibinfo{author}{Shytov, A.~V.}, \bibinfo{author}{Rudner, M.~S.} \&
  \bibinfo{author}{Levitov, L.~S.}
\newblock \bibinfo{title}{Klein backscattering and {F}abry-{P}erot interference
  in graphene heterojunctions}.
\newblock \emph{\bibinfo{journal}{Phys. Rev. Lett.}}
  \textbf{\bibinfo{volume}{101}}, \bibinfo{pages}{156804}
  (\bibinfo{year}{2008}).

\bibitem{Mayorov1}
\bibinfo{author}{Mayorov, A.~S.} \emph{et~al.}
\newblock \bibinfo{title}{Micrometer-scale ballistic transport in encapsulated
  graphene at room temperature}.
\newblock \emph{\bibinfo{journal}{Nano Lett.}} \textbf{\bibinfo{volume}{11}},
  \bibinfo{pages}{2396--2399} (\bibinfo{year}{2011}).

\bibitem{Williams1}
\bibinfo{author}{Williams, J.~R.}, \bibinfo{author}{Low, T.},
  \bibinfo{author}{Lundstrom, M.~S.} \& \bibinfo{author}{Marcus, C.~M.}
\newblock \bibinfo{title}{Gate-controlled guiding of electrons in graphene}.
\newblock \emph{\bibinfo{journal}{Nat. Nanotech.}}
  \textbf{\bibinfo{volume}{6}}, \bibinfo{pages}{222--225}
  (\bibinfo{year}{2011}).

\bibitem{Whisper1}
\bibinfo{author}{Zhao, Y.} \emph{et~al.}
\newblock \bibinfo{title}{Creating and probing electron whispering-gallery
  modes in graphene}.
\newblock \emph{\bibinfo{journal}{Science}} \textbf{\bibinfo{volume}{348}},
  \bibinfo{pages}{672--675} (\bibinfo{year}{2015}).

\bibitem{Gu1}
\bibinfo{author}{Gu, N.}, \bibinfo{author}{Rudner, M.} \&
  \bibinfo{author}{Levitov, L.}
\newblock \bibinfo{title}{Chirality-assisted electronic cloaking of confined
  states in bilayer graphene}.
\newblock \emph{\bibinfo{journal}{Phys. Rev. Lett.}}
  \textbf{\bibinfo{volume}{107}}, \bibinfo{pages}{156603}
  (\bibinfo{year}{2011}).

\bibitem{Liang1}
\bibinfo{author}{Liang, W.} \emph{et~al.}
\newblock \bibinfo{title}{Fabry-perot interference in a nanotube electron
  waveguide}.
\newblock \emph{\bibinfo{journal}{Nature}} \textbf{\bibinfo{volume}{411}},
  \bibinfo{pages}{665--669} (\bibinfo{year}{2001}).

\bibitem{Wu1}
\bibinfo{author}{Wu, Y.} \emph{et~al.}
\newblock \bibinfo{title}{Quantum behavior of graphene transistors near the
  scaling limit}.
\newblock \emph{\bibinfo{journal}{Nano Lett.}} \textbf{\bibinfo{volume}{12}},
  \bibinfo{pages}{1417--1423} (\bibinfo{year}{2012}).

\bibitem{MAllen}
\bibinfo{author}{Allen, M.} \emph{et~al.}
\newblock \bibinfo{title}{Spatially resolved edge currents and guided-wave
  electronic states in graphene}  (\bibinfo{year}{2015}).
\newblock \eprint{http://arxiv.org/abs/1504.07630}.

\bibitem{Nishio1}
\bibinfo{author}{Nishio, T.} \emph{et~al.}
\newblock \bibinfo{title}{Supercurrent through inas nanowires with highly
  transparent superconducting contacts}.
\newblock \emph{\bibinfo{journal}{Nanotechnology}}
  \textbf{\bibinfo{volume}{22}}, \bibinfo{pages}{445701}
  (\bibinfo{year}{2011}).

\bibitem{Jorgensen1}
\bibinfo{author}{Jorgensen, H.~I.}, \bibinfo{author}{Grove-Rasmussen, K.},
  \bibinfo{author}{Novotny, T.}, \bibinfo{author}{Flensberg, K.} \&
  \bibinfo{author}{Lindelof, P.~E.}
\newblock \bibinfo{title}{Electron transport in single-wall carbon nanotube
  weak links in the fabry-perot regime}.
\newblock \emph{\bibinfo{journal}{Phys. Rev. Lett.}}
  \textbf{\bibinfo{volume}{96}}, \bibinfo{pages}{207003}
  (\bibinfo{year}{2006}).

\bibitem{DelftFP}
\bibinfo{author}{Calado, V.~E.} \emph{et~al.}
\newblock \bibinfo{title}{Ballistic josephson junctions in edge-contacted
  graphene}  (\bibinfo{year}{2015}).
\newblock \eprint{http://arxiv.org/abs/1501.06817}.

\bibitem{GeimFP}
\bibinfo{author}{Shalom, M.~B.} \emph{et~al.}
\newblock \bibinfo{title}{Proximity superconductivity in ballistic graphene,
  from fabry-perot oscillations to random andreev states in magnetic field}
  (\bibinfo{year}{2015}).
\newblock \eprint{http://arxiv.org/abs/1504.03286}.

\bibitem{Beenakker1}
\bibinfo{author}{Beenakker, C. W.~J.}
\newblock \bibinfo{title}{Andreev reflection and {K}lein tunneling in
  graphene}.
\newblock \emph{\bibinfo{journal}{Rev. Mod. Phys.}}
  \textbf{\bibinfo{volume}{80}}, \bibinfo{pages}{1337--1354}
  (\bibinfo{year}{2008}).

\bibitem{Blake1}
\bibinfo{author}{Blake, P.} \emph{et~al.}
\newblock \bibinfo{title}{Influence of metal contacts and charge inhomogeneity
  on transport properties of graphene near the neutrality point}.
\newblock \emph{\bibinfo{journal}{Solid State Comm.}}
  \textbf{\bibinfo{volume}{149}}, \bibinfo{pages}{1068--1071}
  (\bibinfo{year}{2009}).

\bibitem{Young1}
\bibinfo{author}{Young, A.~F.} \& \bibinfo{author}{Kim, P.}
\newblock \bibinfo{title}{Quantum interference and klein tunneling graphene
  heterojunctions}.
\newblock \emph{\bibinfo{journal}{Nature Physics}}
  \textbf{\bibinfo{volume}{5}}, \bibinfo{pages}{222--226}
  (\bibinfo{year}{2009}).

\bibitem{Grushina1}
\bibinfo{author}{Grushina, A.~L.}, \bibinfo{author}{Ki, D.} \&
  \bibinfo{author}{Morpurgo, A.}
\newblock \bibinfo{title}{A ballistic pn junction in suspended graphene with
  split bottom gates}.
\newblock \emph{\bibinfo{journal}{App. Phys. Lett.}}
  \textbf{\bibinfo{volume}{102}}, \bibinfo{pages}{223102}
  (\bibinfo{year}{2013}).

\bibitem{Rickhaus1}
\bibinfo{author}{Rickhaus, P.} \emph{et~al.}
\newblock \bibinfo{title}{Ballistic interferences in suspended graphene}
  (\bibinfo{year}{2013}).
\newblock \eprint{http://arxiv.org/abs/1304.6590}.

\bibitem{Allen1}
\bibinfo{author}{Allen, M.~T.}, \bibinfo{author}{Martin, J.} \&
  \bibinfo{author}{Yacoby, A.}
\newblock \bibinfo{title}{Gate-defined quantum confinement in suspended bilayer
  graphene}.
\newblock \emph{\bibinfo{journal}{Nat. Commun.}} \textbf{\bibinfo{volume}{3}},
  \bibinfo{pages}{934} (\bibinfo{year}{2012}).

\bibitem{Rittenhouse1}
\bibinfo{author}{Rittenhouse, G.~E.} \& \bibinfo{author}{Graybeal, J.~M.}
\newblock \bibinfo{title}{Fabry-perot interference peaks in the critical
  current for ballistic superconductor-normal-metal-superconductor {J}osephson
  junctions}.
\newblock \emph{\bibinfo{journal}{Phys. Rev. B}} \textbf{\bibinfo{volume}{49}},
  \bibinfo{pages}{1182--1187} (\bibinfo{year}{1994}).

\bibitem{PJH1}
\bibinfo{author}{Jarillo-Herrero, P.}, \bibinfo{author}{van Dam, J.~A.} \&
  \bibinfo{author}{Kouwenhoven, L.~P.}
\newblock \bibinfo{title}{Quantum supercurrent transistors in carbon
  nanotubes}.
\newblock \emph{\bibinfo{journal}{Nature}} \textbf{\bibinfo{volume}{439}},
  \bibinfo{pages}{953--956} (\bibinfo{year}{2006}).

\bibitem{TinkhamBook}
\bibinfo{author}{Tinkham, M.}
\newblock \emph{\bibinfo{title}{Introduction to Superconductivity}}
  (\bibinfo{publisher}{McGraw-Hill Book Co.}, \bibinfo{address}{New York, NY},
  \bibinfo{year}{1975}).

\bibitem{DasSarma1}
\bibinfo{author}{Hui, H.~Y.}, \bibinfo{author}{Lobos, A.~M.},
  \bibinfo{author}{Sau, J.~D.} \& \bibinfo{author}{Sarma, S.~D.}
\newblock \bibinfo{title}{Proximity-induced superconductivity and {J}osephson
  critical current in quantum spin {H}all systems}  (\bibinfo{year}{2014}).
\newblock \eprint{http://arxiv.org/abs/1410.4205}.

\bibitem{Averin95}
\bibinfo{author}{Averin, D.} \& \bibinfo{author}{Bardas, A.}
\newblock \bibinfo{title}{ac josephson effect in a single quantum channel}.
\newblock \emph{\bibinfo{journal}{Phys. Rev. Lett.}}
  \textbf{\bibinfo{volume}{75}}, \bibinfo{pages}{1831--1834}
  (\bibinfo{year}{1995}).

\end{thebibliography}
\end{document}


\bibliographystyle{nature}

\begin{center}
\huge{{Supplementary Information for}}\\[0.5 cm]
\Large{\textbf{Visualization of phase-coherent electron interference in a ballistic graphene Josephson junction}}\\[0.5 cm]
\large{M. T. Allen, O. Shtanko, I. C. Fulga, J. I.-J. Wang, D. Nurgaliev, K. Watanabe, T. Taniguchi, A. R. Akhmerov, P. Jarillo-Herrero, L. S. Levitov, and A. Yacoby$^{*}$} \\[0.5 cm]

$^{*}$Correspondence to: yacoby@physics.harvard.edu\\[1.0 cm] 
\end{center}








\textbf{Supplementary Materials and Methods}\\

\underline{Fabrication and design of ballistic graphene Josephson junctions}\\

\textit{Suspended Josephson junctions}:  We investigate suspended Josephson junctions of two types.

The first type, which corresponds to data shown in Fig. 1b-d and Supplementary Fig. S1, features superconducting electrodes in the interior of the flake.  Graphene is mechanically exfoliated directly onto on a 300 nm SiO$_2$ dielectric layer that coats a doped silicon wafer serving as a global back gate.  Next, thin Cr/Au leads are defined using e-beam lithography in a pseudo-four probe geometry in order to make electrical contact to the bilayer graphene device.  These contacts are spaced roughly 1-1.5 $\mu$m apart in order to leave room for the superconducting contacts that will eventually define the Josephson junction itself.  These Cr/Au (3/30 nm) contacts are deposited using thermal evaporation, followed by immersion in acetone for metal liftoff.  Next, thick gold electrodes are defined in a way that overlaps the outer edges of the thin contacts, thus maintaining electrical contact to the flake. The thick electrodes serve a dual purpose: (1) to provide structural support and mechanically hold up the entire suspended graphene Josephson junction and (2) provide an electrical connection between the Josephson junction and the bondpads.  After an evaporation mask is defined using e-beam lithography, Cr/Au (3/200 nm) is deposited. To define the Josephson junction, a pair of rectangular Ti/Al superconducting contacts are patterned in the interior of the flake and extending over the thin Cr/Au leads to maintain electrical contact to the bondpads.  The superconducting electrodes are patterned using e-beam lithography, followed by thermal evaporation of a 10 nm Ti adhesion layer and a 70 nm superconducting Al layer.    Finally, in order to protect the superconductor from degradation in acid during the suspension process, a PMMA polymer etch mask is defined over the superconducting contacts using e-beam lithography. After development, the entire chip is immersed in a buffered oxide wet etchant to remove 150 nm of the underlying SiO$_2$ dielectric layer, leaving the Josephson junction fully suspended.  Immediately following the etch, the substrate is immersed in methanol, followed by an acetone soak to dissolve the PMMA mask, after which the chip is again immersed in methanol and dried in a critical point dryer.\\

The second type of suspended Josephson junction, which corresponds to the data in Fig. 2-3, features superconducting electrodes that extend over the full width of the flake.  This Josephson geometry is preferable for imaging current flow due to the uniform distance between contacts and rectangular junction dimensions.  Devices are fabricated on a 300 nm SiO$_2$ dielectric layer that coats a doped silicon wafer that serves as a global back gate.  Bilayer graphene flakes are deposited over predefined narrow trenches that are etched into the SiO$_2$ with a depth of $150$ nm.  Next, thin Cr/Au contacts and bondpads are defined using e-beam lithography in a pseudo-four probe geometry in order to make electrical contact to the bilayer graphene device.  These contacts are spaced far apart on either side of the etched trench in order to leave room for the superconducting contacts that will eventually define the Josephson junction itself.  These Cr/Au (3/30 nm) contacts are deposited using thermal evaporation.  The devices are then immersed in acetone for metal liftoff, transferred immediately into methanol, and carefully dried using a critical point dryer due the delicate nature of suspended graphene membranes.  To construct the Josephson junction, superconducting Ti/Al contacts are patterned along the trench edges using e-beam lithography and with width large enough to achieve electrical contact with the Cr/Au leads.  The superconducting contacts are deposited using thermal evaporation with the following procedure: a 10 nm Ti adhesion layer is deposited, followed by a 50 nm layer of Al superconductor. As with the previous step, metal liftoff is conducted by immersion in acetone and methanol, followed by drying in a critical point dryer.  The motivation for using Cr/Au bondpads is to achieve the best possible electrical connection to the gold bonding wires and sample holder pins.  Aluminum, by contrast, oxidizes upon exposure to air and forms intermetallic compounds at the interface with gold bonding wire, which would be expected to degrade electrical contact.  Devices are current annealed in vacuum at dilution refrigerator temperatures in order to remove organic processing residues and enhance quality. All low temperature data is collected using standard lockin measurement techniques in a Leiden Cryogenics Model Minikelvin 126-TOF dilution refrigerator with a base temperature of $\sim 10$ mK.\\

\textit{Suspended Josephson junction device dimensions}:

Sample \textit{B1}: Fig. 1b-d, Supplementary Fig. S1: main panel plot, right insets; blue curve in the left inset. Type 1 geometry. Distance between superconducting electrodes: 500 nm.  Width of superconducting contacts (defines transverse dimension of junction): 1.7 $\mu$m.

Sample \textit{B2}: Fig. 2a-b, Supplementary Fig. S3: Type 2 geometry. Distance between superconducting electrodes: 350 nm.  Junction width: 1.7 $\mu$m.

Sample \textit{B3}: Fig. 3, Supplementary Fig. S6: Type 2 geometry. Distance between superconducting electrodes: 350 nm.  Junction width: 1.7 $\mu$m.  Note: Data sets \textit{B2} and \textit{B3} are from the same physical device but are collected after different current annealing iterations and thus have different disorder configurations.

Sample \textit{B4}: Supplementary Fig. S1: green curve in the left inset. Type 1 geometry. Distance between superconducting electrodes: 500 nm.  Width of superconducting contacts (defines transverse dimension of junction): 3.2 $\mu$m.

Sample \textit{B5}: Supplementary Fig. S1: red curve in the left inset. Type 1 geometry. Distance between superconducting electrodes: 500 nm.  Width of superconducting contacts: 1.65 $\mu$m.

Sample \textit{B6}: Supplementary Fig. S4-S5: Type 2 geometry. Distance between superconducting electrodes: 350 nm.  Junction width: 1.5 $\mu$m.

Sample \textit{B7}: Supplementary Fig. S4: Type 2 geometry. Distance between superconducting electrodes: 350 nm.  Junction width: 1.5 $\mu$m. Note: Data sets \textit{B6} and \textit{B7} are from the same physical device but are collected after different current annealing iterations and thus have different disorder configurations.\\

\textit{Josephson junctions on hBN}:
To investigate a separate device design, the gate-defined FP cavity, we also consider one dual-gated monolayer graphene Josephson junction encapsulated in hexangonal boron nitride (hBN).  By isolating the graphene from the surface roughness and charge disorder associated with the underlying silicon dioxide gate dielectric, hBN substrates enable high device quality to be achieved, which is a crucial ingredient for observing ballistic charge transport.  This Josephson junction has a distance of 750 nm between the superconducting contacts and a flake width of 2.7 $\mu$m.  The top gate length, which defines the size of the FP resonator, is $\sim$100 nm.  The superconducting electrodes consist of an adhesion layer of Ti (10 nm) and a superconducting layer of Al (60 nm).  The top gate consists of Ti/Au (5/50 nm).  The thicknesses of the top and bottom hBN flakes that encapsulate the graphene are $\sim$19 nm and $\sim$30 nm, respectively, as measured by atomic force microscopy (AFM).\\

\textit{Encapsulated Josephson junction device dimensions}:

Sample \textit{M1}: Fig. 1e, Supplementary Fig. S2: Distance between superconducting electrodes: 750 nm.  Junction width: 2.7 $\mu$m.  The top gate length: $\sim$100 nm. \\

\underline{Fourier method for extraction of supercurrent density distribution}\\

In order to disentangle edge from bulk current flow through the resonator, we employ the Fourier techniques of Dynes and Fulton to reconstruct the real-space supercurrent distribution from the  magnetic interference pattern $I_c(B)$.  This procedure, described in detail in Ref. (37), is briefly summarized here.  When a magnetic field $B$ is applied perpendicular to the junction area, the critical current $I_c(B)$ through a Josephson  junction is:
\begin{equation}
I_c(B)=|\mathcal{I}_c(B)|=\left| \int_{-\infty}^{\infty} J(x) \exp(2\pi i(L+l_{Al})Bx/\Phi_0) dx \right|
\end{equation}
where $x$ is the dimension along the width of the superconducting contacts (labeled in Fig. 1), $L$ is the distance between contacts, $l_{Al}$ is the magnetic penetration length scale (determined by the London penetration depth of the superconductor and flux focusing), and $\Phi_0=h/2e$ is the flux quantum.  This integral expression applies in the narrow junction limit where $L \ll W$, relevant for our system. \\

Observing that $\mathcal{I}_c(B)$ represents the complex Fourier transform of the current density distribution $J(x)$, one can apply Fourier methods to extract the spatial structure of current-carrying electronic states.  Because the antisymmetric component of $J(x)$ vanishes in the middle of the junction, the relevant quantity for analyzing edge versus bulk behavior is the symmetric component of distribution.  By reversing the sign of $I_c(B)$ for alternating lobes of the superconducting interference patterns, we reconstruct $\mathcal{I}_c(B)$ from the recorded critical current.  One can determine the real-space current density distribution across the sample by computing the inverse Fourier transform:
\begin{equation}
J_s(x)\approx\int_{-\infty}^{\infty} \mathcal{I}_c(B)
\exp(2\pi i(L+l_{Al})Bx/\Phi_0) dB
\end{equation}

We employ a raised cosine filter to taper the window at the endpoints of the scan in order to reduce convolution artifacts due to the finite scan range $B_{min}<B<B_{max}$.  This the explicit expression used is:
\begin{equation}
J_s(x)\approx\int_{B_{min}}^{B_{max}} \mathcal{I}_c(B) 
\cos^n(\pi B /2 L_B)
\exp(2\pi i(L+l_{Al})Bx/\Phi_0) dB
\end{equation}
where $n=0.5-1$ and $L_B=(B_{max}-B_{min})/2$ is the magnetic field range of the scan.\\

\newpage

\textbf{Supplementary Figures}
\begin{center}
 \begin{figure}[h!]
 \includegraphics[width=1.0\textwidth]{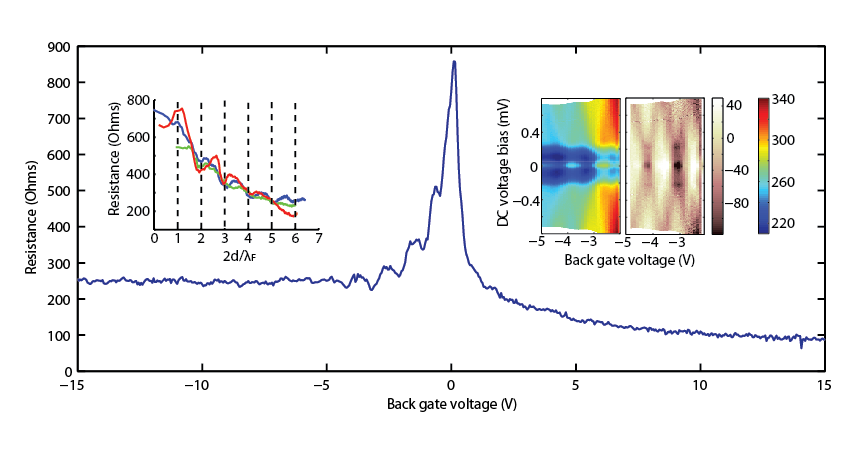}
	  \caption{\textbf{Characterization of Fabry-P\'erot interference conditions in ballistic graphene Josephson junctions}. Main panel: Plot of the normal resistance $R_n$, obtained by sweeping the gate voltage $V_b$ at a fixed bias exceeding $I_c$ (repeated here from Fig. 1b for reference to the other panels). Data was collected from sample \textit{B1}. \textit{Right inset:} Fabry-P\'erot diamonds obtained using voltage bias spectroscopy, as shown in color maps of $R(\Omega)$ and its derivative $dR_n/dV_b$, as function of back gate voltage $V_b$ and voltage bias $V_{DC}$. Data from sample \textit{B1}.  \textit{Left inset:} $R_n$ plotted versus $2d/\lambda_F$, where $d$ is the effective junction length and $\lambda_F$ is the Fermi wavelength.  By comparing the junction length $L$ to the effective size $d$ extracted from fits, we determined that the contact-doped regions extend at most 100 nm into the channel, consistent with the results of scanning photocurrent studies. Reproducibility of the oscillation period is demonstrated in three devices of length $L=500$ nm.  The blue resistance curve is from sample \textit{B1}, the green curve is from sample \textit{B4}, and the red curve is from sample \textit{B5} and offset by -250 $\Omega$. Resonances marked by dips in resistance appear when constructive interferences conditions are satisfied. }
 \end{figure}
\end{center}

\begin{center}
 \begin{figure}[h!]
 \includegraphics[width=0.8\textwidth]{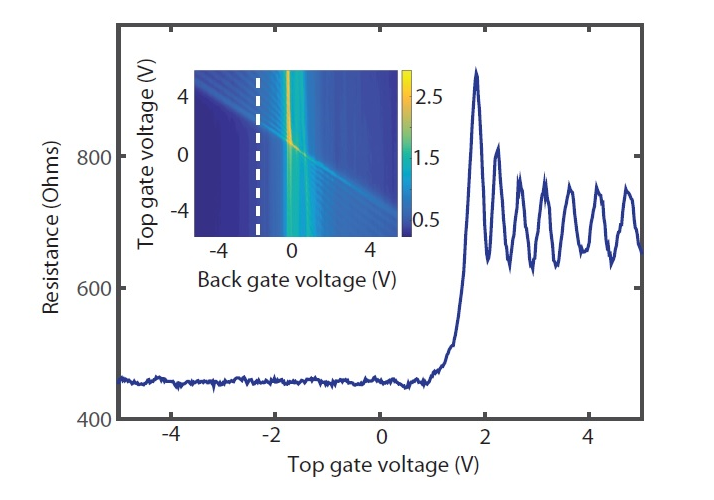}
	  \caption{\textbf{Ballistic resistance oscillations in a gate-defined Fabry-P\'erot interferometer on hBN. } Normal state resistance of a monolayer graphene device on hBN as a function of top gate when the back gate is held fixed at -1.75 V (resistance line cut corresponds to the white dotted line in the inset).  Data was collected from sample \textit{M1}. The oscillation period agrees with a Fabry-P\'erot model with a cavity length $~\sim 100$ nm. Inset shows that oscillation occurs in p-n-p and n-p-n regions, which is characteristic of Klein tunneling in monolayer graphene. The oscillations also suggest the ballistic nature of electronic transport in the locally gated region.}
 \end{figure}
\end{center}

\begin{center}
 \begin{figure}[h!]
 \includegraphics[width=1.0\textwidth]{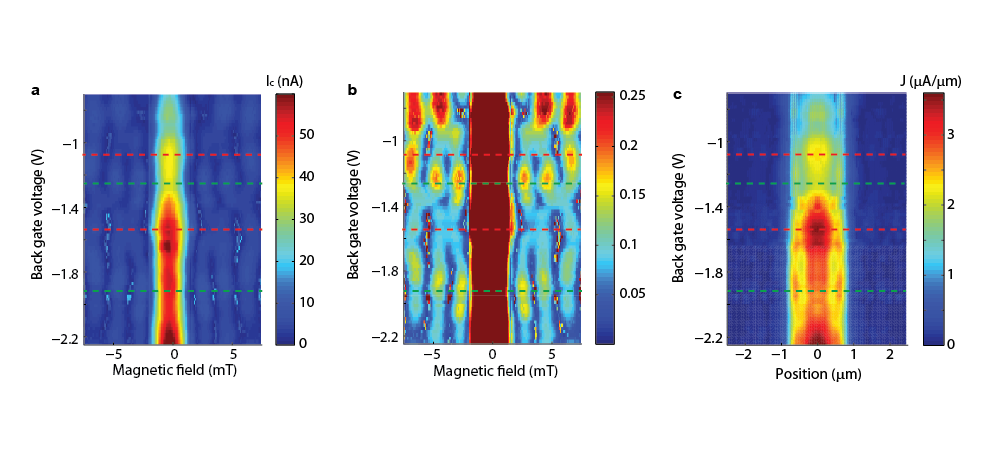}
	  \caption{\textbf{Nontrivial current flow through a graphene Fabry-P\'erot resonator, as revealed by Fraunhofer interferometry.}  \textbf{(a)} Unsaturated map of the critical current $I_c(B)$ data from Fig. 2a, plotted over a full color scale range. \textbf{(b)} Plot of normalized critical current $I_c(B)/I_c(B=0)$ from the data in Fig. 2a, indicating a nontrivial dependence of Fraunhofer interference on cavity resonances.  Red and green dotted lines indicate \textit{on} and \textit{off} resonance conditions for the cavity, respectively. \textbf{(c)} Real-space supercurrent density distribution $J(x)$ extracted from the Fraunhofer interference $I_c(B)$ data in Fig. 2a using Fourier techniques.  (Fig. 2b in the main text is a plot of the real-space normalized supercurrent density distribution for this data set.)  Data was collected from sample \textit{B2}.}
 \end{figure}
\end{center}

\begin{center}
 \begin{figure}[h!]
 \includegraphics[width=0.85\textwidth]{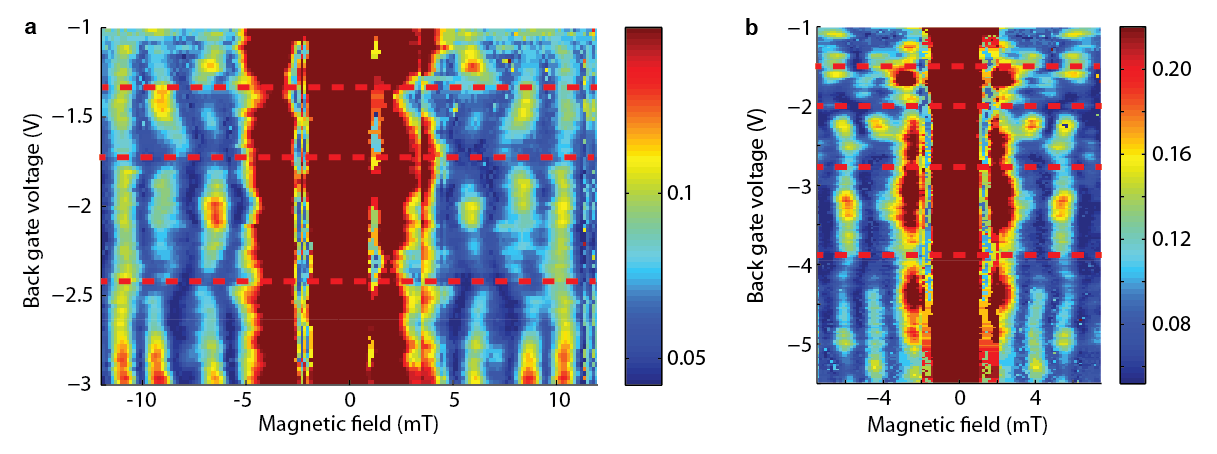}
	  \caption{\textbf{Dependence of normalized Fraunhofer interference on cavity resonances in additional samples.} Plot of normalized critical current $I_c(B)/I_c(B=0)$, indicating nontrivial dependence of Fraunhofer interference on cavity resonances.  Red dotted lines indicate \textit{on} resonance conditions for the cavity.  Data was collected from samples \textit{B6} (panel \textbf{(a)}) and \textit{B7} (panel \textbf{(b)}), which exhibit qualitatively similar behavior to sample \textit{B2} in Supplementary Fig. S3.}
 \end{figure}
\end{center}

\begin{center}
 \begin{figure}[h!]
 \includegraphics[width=1.0\textwidth]{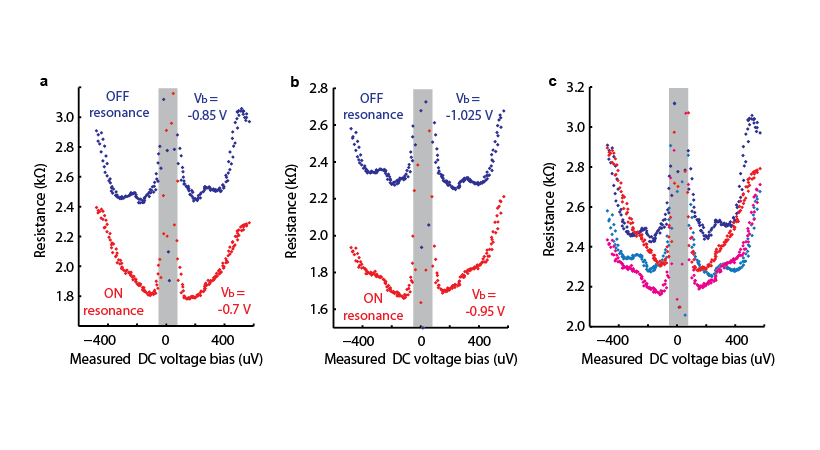}
	  \caption{\textbf{Interplay between multiple Andreev reflection and cavity resonances in an additional device.} Experimental resistance profiles obtained using voltage bias spectroscopy, measured at fixed back gate voltages $V_b$ tuned to \textit{on} or \textit{off} resonance conditions for the cavity (red or blue curves, respectively).  The amplitude of the multiple Andreev reflections, manifested in resistance dips at $2\Delta/n$ for integer $n$, is strongly modulated by cavity transmission and thus exhibits suppression when carrier density is tuned on-resonance.   \textbf{(a)} Red curve: $V_b=-0.7$ V, (on-resonance, corresponding to a dip in normal state resistance). Blue curve: $V_b=-0.85$ V, (off-resonance, corresponding to a peak in normal state resistance). \textbf{(b)} Red curve: $V_b=-0.95$ V (on-resonance);  Blue curve: $V_b=-1.025$ V (off-resonance). In panels (a) and (b), the red curves are offset by $-500 \Omega$ for clarity. \textbf{(c)} All curves from panels (a) and (b), plotted on the same resistance scale. Red curve: $V_b=-0.7$ V (on-resonance); dark blue curve: $V_b=-0.85$ V (off-resonance); magenta curve: $V_b=-0.95$ V (on-resonance).; light blue curve: $V_b=-1.025$ V (off-resonance).  Data was collected from sample \textit{B6}, which shows qualitatively equivalent behavior to sample \textit{B3} in Fig. 3 of the main text.}
 \end{figure}
\end{center}

\begin{center}
 \begin{figure}[h!]
 \includegraphics[width=0.4\textwidth]{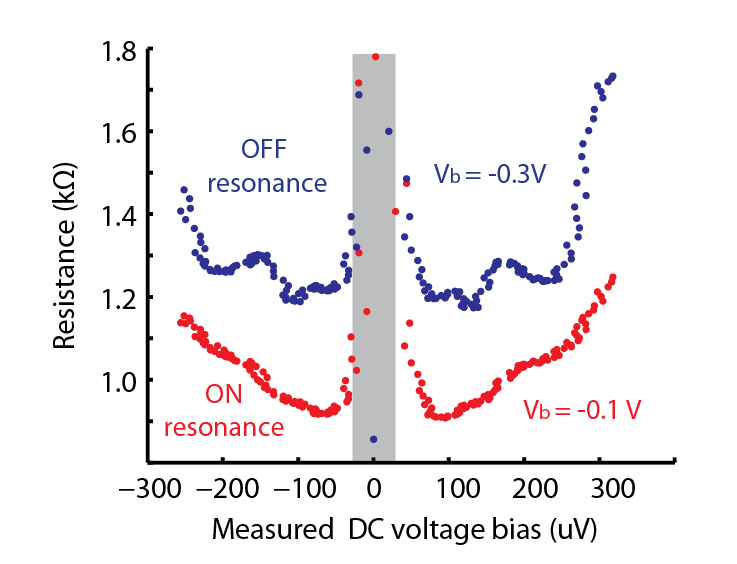}
	  \caption{\textbf{Additional multiple Andreev reflection data sets from sample \textit{B3}.} More voltage bias spectroscopy data from sample in Fig. 3, taken at additional back gate voltages.  Red curve: $V_b=-0.1$ V (on-resonance);  Blue curve: $V_b=-0.3$ V (off-resonance).}
 \end{figure}
\end{center}

\begin{center}
 \begin{figure}[h!]
 \includegraphics[width=1.0\textwidth]{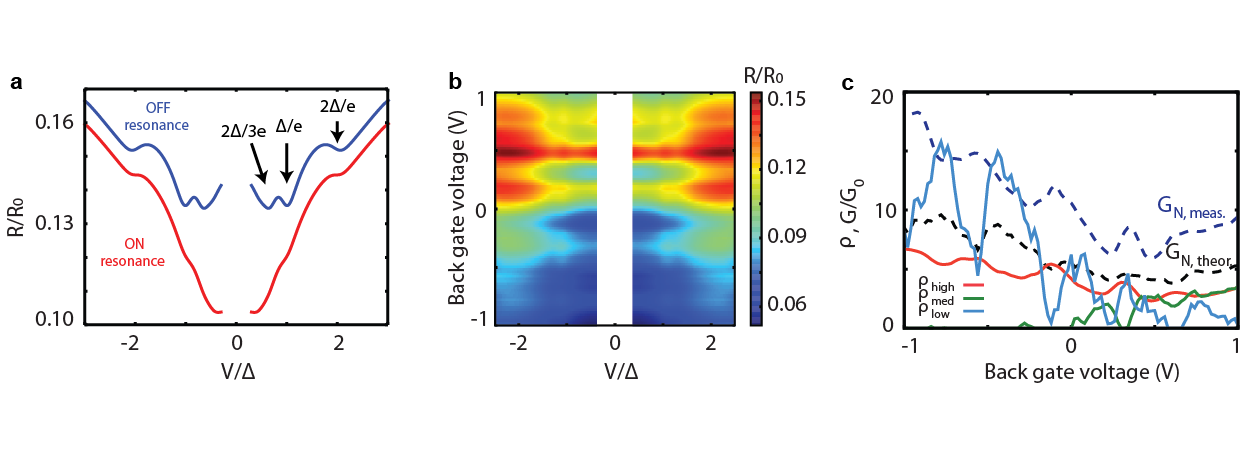}
	  \caption{\textbf{Theoretical dependence of multiple Andreev reflection amplitude on cavity transmission.} \textbf{(a)}  Theoretical resistance profiles as a function of applied bias voltage, corresponding to the experimental data in Fig. 3c. The red curve (on resonance, $V_b=0.3 V$) shows suppressed MAR features, while in the blue curve (off resonance, $V_b=0.5 V$) MAR peaks appear at bias voltages $2\Delta$, $\Delta$, and $2\Delta/3$. The red curve has been shifted upwards by $0.015 R_0$ for clarity.  \textbf{(b)}  Simulated resistance map obtained by fitting the measured data to the short junction model, plotted continuously as a function of applied DC bias voltage $V$ and back gate voltage. The theoretical resistance profile is in good agreement with the experimental one (Fig. 3b), showing well defined MAR peaks when the system is off resonance, and suppressed MAR features on resonance.  \textbf{(c)} Mode contributions $\rho$ (thick solid lines) corresponding to large (red), medium (green), and small (blue) transmissions, as a function of back gate voltage. The black and blue dashed lines show the normal state conductance values from measurement and theory, respectively.}
 \end{figure}
\end{center}